 \documentclass[11pt]{article}
 \usepackage[top=1in,bottom=1in,left=1.5in,right=1.5in]{geometry}
  \usepackage{amsmath,amssymb}
  \usepackage{latexsym}
 \usepackage[dvips]{pstricks} 
  \usepackage{pst-node}
  \usepackage[dvips]{graphicx}

  \newcommand{\C}{\mathbb{C}}
  \newcommand{\F}{\mathbb{F}}

  \newcommand{\R}{\mathbb{R}}
  
  \renewcommand{\a}{\mathbf{a}}

  \newcommand{\bt}{\mathbf{t}}
  
  \newcommand{\U}{\mathbf{U}}
  \renewcommand{\u}{\mathbf{u}}
  \renewcommand{\v}{\mathbf{v}}
  \newcommand{\V}{\mathbf{V}}

  \newcommand{\x}{\mathbf{x}}
  
  \newcommand{\y}{\mathbf{y}}

  \newcommand{\0}{\mathbf{0}}

  \newcommand{\lam}{\mbox{\boldmath{$\lambda$}}}
  \newcommand{\bA}{\mathbf{A}}

  \newcommand{\cA}{\mathcal{A}}
  \newcommand{\cB}{\mathcal{B}}

  \newcommand{\cH}{\mathcal{H}}

  \newcommand{\cS}{\mathcal{S}}

  \newcommand{\rH}{\mathrm{H}}

  \newcommand{\lan}{\langle}
  \newcommand{\ran}{\rangle}
  \newcommand{\an}[1]{\lan#1\ran}
  
  \newcommand{\hs}{\hspace*{\parindent}}
  \newcommand{\proof}{\hs \textbf{Proof.\ }}
  
  \newcommand{\tr}{\mathop{\mathrm{tr}}\nolimits}

  \newcommand{\trans}{^\top}
  \newcommand{\qed}{\hspace*{\fill} $\Box$\\}

  \newcommand{\rF}{\mathrm{F}}
  
  \newcommand{\rO}{\mathrm{O}}

  \newcommand{\sig}{\boldsymbol{\sigma}}
  
  \newcommand{\et}{\boldsymbol{\eta}}

  \newcommand{\rank}{\mathrm{rank\;}}

  \newtheorem{theo}{\bfseries \hs Theorem}[section]
  
  \newtheorem{prop}[theo]{\bfseries \hs Proposition}
  
  \newtheorem{lemma}[theo]{\bfseries \hs Lemma}
  \newtheorem{corol}[theo]{\bfseries \hs Corollary}

  \numberwithin{equation}{section} 

 \renewcommand{\span}{\mathrm{span}}

\begin{document}

 \title{Additive invariants on quantum channels\\and
 applications to regularized minimum entropy\thanks
 {This research started during author's participation in AIM workshop
 ``Geometry and representation theory of tensors for computer science,
 statistics and other areas", July 21-25, 2008.}}
 \author{Shmuel Friedland\\
 Department of Mathematics, Statistics, and Computer Science\\
  University of Illinois at Chicago\\
  Chicago, Illinois 60607-7045, USA\\
  \texttt{E-mail:
  friedlan@uic.edu}}

\date{February 9, 2009 }
 \maketitle
 \begin{abstract}
 We introduce two additive invariants of output quantum channels.
 If the value of one these invariants is less than $1$
 then the logarithm of the inverse of its value
 is a positive lower bound for the regularized minimum entropy
 of an output quantum channel.
 We give a few examples in which
 one of these invariants is less than $1$.
 We also study the special cases where the above both invariants
 are equal to $1$.
 \end{abstract}



 \section{Introduction}\label{intro}
 Denote by $\cS_n(\C)$ the Hilbert space of $n\times n$ hermitian
 matrices, where $\an{X,Y}=\tr XY$.  Denote by $\cS_{n,+,1}(\C)\subset\cS_{n,+}(\C)
 \subset \cS_n(\C)$
 the convex set of positive hermitian matrices of trace one, and the cone of
 positive hermitian matrices respectively.
 A quantum channel is a
 completely positive linear transformation
 $\tau:\cS_n(\C)\to\cS_m(\C)$:
 \begin{equation}\label{defcomposop}
 \tau(X)=\sum_{i=1}^l A_i X A_i^*, \quad A_1,\ldots,A_l\in
 \C^{m\times n}, \;X\in \cS_n(\C),
 \end{equation}
 which is trace preserving:
 \begin{equation}\label{tracecon}
 \sum_{i=1}^l A_i^*A_i=I_n,
 \end{equation}
 The  \emph{minimum} entropy output of a quantum channel $\tau$ is
 defined
 \begin{equation}\label{defminentrop}
 \rH(\tau)=
 \min_{X\in \cS_{n,+,1}(\C)} -\tr \tau(X)\log \tau(X).
 \end{equation}
 If $\eta:\cS_{n'}(\C)\to \cS_{m'}(\C)$ is another quantum channel,
 then it is well known $\tau\otimes\eta$ is a quantum channel, and
 \begin{equation}\label{subadit}
 \rH(\tau\otimes \eta)\le \rH(\tau)+\rH(\eta).
 \end{equation}
 Hence the sequence $\rH(\otimes^p \tau), p=1,\ldots,$ is
 subadditive.  Thus the following limit exists:
 \begin{equation}\label{defregminent}
 \rH_r(\tau)=\lim_{p\to\infty} \frac{\rH(\otimes^p\tau)}{p},
 \end{equation}
 and is called the \emph{regularized} minimum entropy of
 quantum channel.  Clearly,  $\rH_r(\tau)\le \rH(\tau)$.

 One of the major open problem of quantum information theory
 is the additivity conjecture, which claims that
 equality holds in (\ref{subadit}).  This additivity conjecture
 has several equivalent forms \cite{Sho03}.  If the additivity
 conjecture holds then $\rH_r(\tau)=\rH(\tau)$, and the
 computation of $\rH_r(\tau)$ is relatively simple.
 There are known cases where
 the additivity conjecture is known, see references in \cite{HW08}.  It is also known
 that the $p$ analog of the additivity conjecture is wrong
 \cite{HW08}.
 It was shown in \cite{AIM08} that the additivity of the
 entanglement of subspaces fails over the real numbers.
 It was recently shown by Hastings \cite{Has08} that
 the additivity conjecture is false.
 Hence the computation of $\rH_r(\tau)$ is hard.  This is the standard situation in
 computing the entropy of Potts models in statistical physics,
 e.g. \cite{FP05}.

 The first major result of this paper gives a nontrivial lower bound on
 $\rH_r(\tau)$ for certain quantum channels.  This is done by
 introducing two additive invariants on quantum channels.
 Let
 \begin{equation}\label{defboldA}
 \bA(\tau):=\sum_{i=1}^l A_i  A_i^*\in \cS_{m,+}(\C).
 \end{equation}
 Then $\log\lambda_1(\bA(\tau))=\log\|\bA(\tau)\|$, where $\lambda_1(\bA)$
 is the maximal eigenvalue of $\bA(\tau)$, is the first
 additive invariant of quantum channels, with respect to
 tensor products.
 Let $\sigma_1(\tau)=\|\tau\|\ge \sigma_2(\tau)\ge\ldots\ge 0 $ be the first and the second singular
 value of the linear transformation given by $\tau$.  Then
 $\log\sigma_1(\tau)$ is the second additive invariant.
 (These two invariants are incomparable in general, see \S5.)
 The main result of this paper is the inequality
 $$\rH_r(\tau)\ge
 \max(-\log\lambda_1(\bA(\tau)),-\log\sigma_1(\tau)).$$
 This inequality is nontrivial only if
 $\min(\lambda_1(\bA(\tau)),\sigma_1(\tau))<1$.
 In \S5 section we give examples where $\min(\lambda_1(\bA(\tau)),\sigma_1(\tau))<1$.
 If $\lambda_1(\bA(\tau))<1$ then the inequality $\rH_r(\tau)\ge
 -\log\lambda_1(\bA(\tau))$ can be improved, see \S4.

 It is easy to show that
 $\lambda_1(\bA(\tau))\ge \frac{n}{m}$ and
 $\sigma_1(\tau)\ge \frac{\sqrt{n}}{\sqrt{m}}$, see Proposition \ref{lowbdl1a}.
 Hence, for $m\le n$ we
 must have $\lambda_1(\bA(\tau)),\sigma_1(\tau)\ge 1$.  Perhaps, the most
 interesting case is the case where $m=n$.  Furthermore,
 very interesting quantum channels $\tau$ are \emph{unitary} quantum
 channels, which are of the form
 (\ref{defcomposop}), where
 \begin{equation}\label{spectau}
 A_i=t_iQ_i,\; Q_iQ^*_i=Q_i^*Q_i=I_n,\; i=1,\ldots,l,\;
 \bt=(t_1,\ldots,t_l)\trans\in\R^l,\bt\trans\bt=1.
 \end{equation}
 In that case $\lambda_1(\bA(\tau))=\sigma_1(\tau)=1$.
 Note the counter example to the additivity conjecture in
 \cite{Has08} is of this form.
 A quantum channel $\tau:\cS_n(\C)\to \cS_m(\C)$ is called
 a \emph{bi}-quantum channel if $m=n$ and $\tau^*:\cS_n(\C)\to
 \cS_n(\C)$ is also a quantum channel.  That is
 $\bA(\tau)=I_n$ and it follows that $\sigma_1(\tau)=1$.
 Note that a unitary quantum channel is a
 bi-quantum channel.
 The second major result of this paper is the lower
 bound
 $$  \rH(\otimes^p\tau)\ge -\frac{1}{2} \log (\sigma_2(\tau)^2 +
 \frac{1-\sigma_2(\tau)^2}{n^p}), \; p=1,\ldots,$$
 for a bi-quantum channel.  Note that this lower is
 nontrivial if $\sigma_2(\tau)<1$.
 We show that the condition $\sigma_2(\tau)<1$ holds for a
 generic unitary channel with $l\ge 3$.

 \section{Preliminary results}\label{prelim}

 Let $\F=\R,\C$ be the field of real and complex numbers respectively,
 and denote by $\F^n$ the vector space of
 the column vectors $\x=(x_1,\ldots,x_n)\trans$ with
 coordinates in $\F$.  We view $\F^n$ as an inner product
 space, i.e. \emph{Hilbert space} $\cH_A$, with the inner product
 $\an{\x,\y}:=\y^*\x=\sum_{j=1}^n  \bar y_j x_j$.

 View $\F^m\otimes \F^n$ as the set of $m\times n$ matrices
 with entries in $\F$, denoted by $\F^{m\times n}$.
 Equivalently, if we identify $\F^m$ with the Hilbert space
 $\cH_B$ then $\F^{m\times n} \thickapprox \cH_B\otimes \cH_A$.
 Recall that on $\F^{m\times n}$ we have the inner product
 $\an{A,B}:=\tr AB^*$, where $B^*=A\trans$ if $B\in \R^{m\times
 n}$ and $B^*=(\bar B)\trans$ if $B\in \C^{m\times n}$.

 Denote by $\cS_n(\F)\subset
 \F^{n\times n}$ the real space of self-adjoint matrices.
 I.e.  $\cS_n(\R)$ is the space of real symmetric matrices, and
 $\cS_n(\C)$ is the space of hermitian matrices.
 Let $X\in \cS_n(\F)$.  Denote by
 $\lam(A)=(\lambda_1(X),\ldots,\lambda_n(X))$ the eigenvalue
 set of $X$, where $\lambda_1(A)\ge \ldots\ge \lambda_n(X)$.
 Then $\u_1,\ldots,\u_n$ is the corresponding orthonormal basis
 of $\F^n$ consisting of eigenvectors of $X$
 $$X\u_i=\lambda_i(X)\u_i,\; \u_i^*\u_j=\delta_{ij}, \quad
 i,j=1,\ldots,n.$$
 Ky-Fan maximal characterization is, e.g. \cite{Fri81a},
 \begin{equation}\label{kyfan}
 \sum_{j=1}^k \lambda_j(X)=\max_{\x_1,\ldots,\x_k\in\C^n,
 \x_p^*\x_q=\delta_{pq}}\sum_{j=1}^k \x_j^* X\x_j=\sum_{j=1}^k
 \tr(X(\x_j\x_j^*)).
 \end{equation}
 As in physics, we call $X\in\cS_n(\F)$ a \emph{positive} hermitian matrix,
 or simply positive,
 and denoted it by $X\ge 0$, if all eigenvalues of $X$ are
 nonnegative.
 Also for $X,Y\in \cS_n(\F)$ we let $Y\ge X$ if
 $Y-X\ge 0$.  Denote by $\cS_{n,+,1}(\F)\subset\cS_{n,+}(\F)\subset \cS_n(\F)$
 the convex set of positive hermitian matrices of trace one, and the cone of
 positive hermitian matrices respectively.

 Let $A\in \F^{m\times n}$.  Then the positive singular values
 of $A$ are the positive eigenvalues of $\sqrt{AA^*}$, which
 are equal to the positive eigenvalues $\sqrt{A^*A}$.
 Let $\sig(A)=(\sigma_1(A),\sigma_2(A),\ldots, \sigma_l(A))\trans$
 be the vector of singular values of $A\in \F^{m\times n}$, where
 $\sigma_1(A)\ge \sigma_2(A)\ge\ldots\ge \sigma_l(A)\ge 0$ are the singular
 values of $A$ arranged in the decreasing order.  We do not fix
 the number of coordinates $l$ in $\sig(A)$, but recall that
 $\sigma_i(A)=0$ if $i>\min(m,n)$.  (So $l\ge\min(m,n)$.)
 There exists an orthonormal bases
 $\u_1,\ldots,\u_n\in\C^n,\v_1,\ldots,\v_m\in\C^m$, called right and left singular vectors
 of $A$, such that
 \begin{eqnarray}\nonumber
 &&A\u_i=\sigma_i(A)\v_i,\;A^*\v_i=\sigma_i(A)\u_i, \;i=1,\rank
 A,\; A\u_i=\0, A^*\v_i=\0 \textrm{ for } i>\rank A,\\
 &&\label{lefrigsv} \u_i^*\u_j=\delta_{ij}, \;i,j=1,\ldots, n,
 \quad \v_p^*\v_q=\delta_{pq}, \; p,q=1,\ldots, m.
 \end{eqnarray}
 Note that the Frobenius norm
 $\|A\|_F:=\sqrt{\an{A,A}}=\sqrt{\tr(AA^*)}$ is equal to
 $\sqrt{\sum_{i=1}^{\rank A}\sigma_i(A)^2}$.

 Assume that $X\in \cS_n(\F)$.  Then the singular values of $X$
 are the absolute values of the eigenvalues of $X$.  In
 particular, for $X\in\cS_{n,+}(\F)$ we have that
 $\sig(X)=\lam(X)$.

 Recall the well known maximal characterization of the sum of the
 first $k$ singular values of $A\in \F^{m\times n}$ \cite[Thm 3.4.1]{HJ99}.
 \begin{equation}\label{maxcharsv}
 \sum_{j=1}^k \sigma_j(A)=\max_{\x_p\in \F^n,\y_q\in \F^m,
 \x_p^*\x_q=\y_p^*\y_q=\delta_{pq}, p,q=1,\ldots,k} \sum_{j=1}^k
 \y_j^*A\x_j,
 \end{equation}
 for $k=1,\ldots,\min(m,n)$.
 Note that $\sigma_1(A)=\|A\|=\max_{\x^*\x=1} \|A\x\|$, where
 $\|A\|$ is the $\ell_2$ norm of $A$.
 A useful observation is
 \begin{equation}\label{triden}
 \y^*A\x=\tr (A(\x\y^*))=\tr (A(\y\x^*)^*)=\tr ((\x\y^*)A).
 \end{equation}

 For any nonnegative vector $\x=(x_1,\ldots,x_n)\trans\in\R_+^n$
 denote by
 $$\rH(\x):=-\sum_{j=1}^n x_i\log x_i.$$
 Let $\Pi_n\subset\R_+^n$ be the set of probability vectors with
 $n$-coordinates.  Then $\rH(\x)$ is the entropy of a
 probability vector $\x$.
 For $X\in \cS_{n,+}(\F)$ we define the von Neumann entropy
 $$\rH(X):=\rH(\lam(X))=-\tr X\log X.$$
 Note that if $X\in \cS_{n,+,1}(\F)$
 then $\rH(X)=0$ if and only if $X$ is a rank one nonnegative definite matrix with
 trace $1$.

 It is well known that if $\phi: \R_+\to \R$ is a convex
 function then
 $$\phi: \cS_{n,+}(\F)\to \R,\quad \phi(X)=\sum_{i=1}^n
 \phi(\lambda_i(X)), \;X\in\cS_{n}(\F)$$
 is a convex function on $\cS_{n,+}(\F)$.  See for example
 \cite{Fri81a}.  This fact is implied by the
 \emph{majorization} relation
 $$\lam(aX+bY)\prec a\lam(X)+b\lam(Y), \quad a,b\in\R_+, X,Y\in
 \cS_n(\F),$$
 which is equivalent to
 $$\sum_{j=1}^i \lambda_j(aX+bY)\le\sum_{j=1}^i
 (a\lambda_j(X)+b\lambda_j(Y)), \quad i=1,\ldots,n-1,$$
 and the trace equality $\tr(aX+bY)=a\tr X +b\tr Y$.  See
 \cite{HPL52, MO79} for good references on majorization.
 In particular, $-\rH(X)=\tr(X\log X)$ is a convex
 function on $\cS_{n,+}(\F)$.

 In what follows it is convenient to identify $\F^{m_1\times n_1}\otimes
 \F^{m_2\times n_2}$ with $\F^{(m_1m_2)\times (n_1 n_2)}$.
 Assume that $X_i=[x_{pq,i}]_{p=q=1}^{m_i,n_i}\in \F^{m_i\times n_i}$ for $i=1,2$.
 Then we identify $X_1\otimes X_2$ with the Kronecker product,
 which is viewed as $(m_1m_2)\times (n_1n_2)$ matrix given as a
 block matrix $[x_{pq,1}X_2]_{p=q=1}^{m_1,n_1}$.  So
 $X_1\otimes X_2$ maps $\F^{n_1n_2}$ to $\F^{m_1m_2}$.
 Identify $\F^{mn}$ with the matrix space $\F^{n\times m}$.
 Then
 \begin{equation}\label{kronprop}
 (X_1\otimes X_2) (Y)=X_2YX_1\trans, \quad Y\in \F^{n_2\times
 n_1}.
 \end{equation}
 Recall the well known fact that $\rank (X_1\otimes X_2)=\rank
 X_1 \rank X_2$.  Furthermore, all positive singular values of
 $X_1\otimes X_2$ are of the form \cite[Thm 4.2.15]{HJ99}
 $$\sigma_i(X_1)\sigma_j(X_2), \quad i=1,\ldots,\rank
 X_1,\;j=1,\ldots,\rank X_2.$$
 In particular
 \begin{eqnarray}\label{tenprodsv}
 \sigma_1(X_1\otimes X_2)=\sigma_1(X_1)\sigma_1(X_2),\\
 \sigma_{\rank X_1\otimes X_2}(X_1\otimes X_2)=\sigma_{\rank X_1}(X_1)
 \sigma_{\rank X_2}(X_2).\nonumber
 \end{eqnarray}
 Hence we have the additivity of the
 entropy formula
 \begin{equation}\label{additivfor}
 \rH(X_1\otimes X_2)=\rH(X_1)+\rH(X_2) \textrm{ for }
 X_i\in\cS_{n_i,+,1}(\F),\;i=1,2,
 \end{equation}

 \section{Main inequalities}
 In this section we view $\cS_n(\C)$ as $\R^{n^2}$.
 The real inner product on $\cS_n(\C)$ is given by
 $\an{X_1,X_2}=\tr(X_1X_2^*)=\tr(X_1X_2)$.
 Let $\phi:\cS_n(\C)\to \cS_m(\C)$ be a linear, (real), transformation.
 We now apply the notions discussed in the previous section.
 The adjoint linear transformation $\phi^*:\cS_m(\C)\to \cS_n(\C)$ is
 given by the identity
 $$\tr(\phi(X) Y)=\tr(X\phi^*(Y)) \textrm{ for all } X\in
 \cS_n(\C), \;Y\in \cS_m(\C).$$
 The positive singular values of $\phi$ are the positive eigenvalues
 of $(\phi \phi^*)^{\frac{1}{2}}$ or of $(\phi^*
 \phi)^{\frac{1}{2}}$:
 \begin{eqnarray*}
 \sigma_i(\phi)=\sigma_i(\phi^*)=(\lambda_i(\phi\phi^*))^{\frac{1}{2}}=
 (\lambda_i(\phi^*\phi))^{\frac{1}{2}} \; i=1,\ldots, \rank \phi,\\
 \sigma_i(\phi)=\sigma_i(\phi^*)=0 \textrm{ for } i>\rank
 \phi.
 \end{eqnarray*}
 We will denote $\sigma_i(\phi)$ by $\sigma_i$ where no
 ambiguity arises.
 Furthermore, there exist orthonormal bases $\{U_1,\ldots, U_{n^2}\}, \{V_1,\ldots,V_{m^2}\}$
 of $\cS_n(\C),\cS_m(\C)$ respectively, such that the following
 conditions hold.
 \begin{eqnarray}\label{singvalcon}
 \phi(U_i)=\sigma_i V_i, \; \phi_i^*(V_i)=\sigma_i U_i,\quad
 \phi(U_i)=0,\;\phi_i^*(V_i)=0 \textrm{ for } i>\rank \phi,\\
 \label{singvalcon1}
 \tr(U_iU_j)=\delta_{ij} \textrm{ for } i,j=1,\ldots,n,\quad
 \tr(V_pV_q)=\delta_{pq} \textrm{ for } p,q=1,\ldots, m.
 \label{orthcon}
 \end{eqnarray}
 Combine the well known expression of $\tr \phi(X)^2=\|\phi(X)\|^2$ in terms
 of singular values and vectors of $\phi$ to deduce
 \begin{equation}\label{SVDnormid}
 \sum_{i=1}^m \lambda_i(\phi(X))^2= \|\phi(X)\|^2=\sum_{i=1}^{\rank \phi} \sigma_i^2 |\tr U_i X|^2
 \textrm{ for any } X\in \cS_n(\C).
 \end{equation}
 If $m=n$ and $\phi$ is self-adjoint, i.e.
 $\tr(\phi(X)Y)=\tr(X\phi(Y))$, then the singular values of
 $\phi$ are the absolute values of the eigenvalues of $\phi$.
 If an addition $\phi$ is positive operator, i.e.
 $\tr(\phi(X)X)\ge 0$, the singular values of $\phi$ are the
 eigenvalues of $\phi$.  In that case in (\ref{singvalcon}) we
 assume that $X_i=Y_i, i=1,\ldots,n$.
 The maximal characterization (\ref{maxcharsv}) is
 \begin{equation}\label{maxchar}
 \sum_{i=1}^k
 \sigma_i(\phi)=\max_{X_1,\ldots,X_k\in\cS_{n}(\C),Y_1,\ldots,Y_k\in\cS_{m}(\C),
 \tr(X_iX_j)=\tr(Y_iY_j)=\delta_{ij}}
 \sum_{i=1}^k\tr(\phi(X_i)Y_i),
 \end{equation}
 for $k=1,\ldots, \min(m,n)$.  If $m=n$ and $\phi$ is
 self-adjoint and positive we assume that $X_i=Y_i$ for
 $i=1,\ldots,k$.
 Note that $\sigma_1(\phi)=\|\phi\|$.

 A linear mapping $\phi:\cS_n(\C)\to \cS_m(\C)$ is called
 \emph{positive preserving} if $\phi(\cS_{n,+}(\C))\subset \cS_{m,+}(\C)$.
 Since $\cS_{n,+}(\C)$ is a self-adjoint cone, it follows
 $\phi$  is positive preserving if and only if $\phi^*$ is
 positive preserving.  In particular, if $\phi$ is positive
 preserving, then the positive operators $\phi\phi^*$ and
 $\phi^*\phi$ are positive and positive preserving operators.
 Assume that $\phi$ is positive preserving.
 The Krein-Rutman theorem cone preserving theorem, e.g. \cite{BP}, imply that
 in (\ref{singvalcon}) we can choose $U_1\in \cS_{n.+}(\C),
 V_1\in\cS_{m,+}(\C)$.  If $\phi$ is \emph{strict positive
 preserving}, i.e. for each $0\ne X\in\cS_{n,+}(\C)$ $\phi(X)$
 has positive eigenvalues, then
 $U_1\in\cS_{n,+}(\C),V_1\in\cS_{m,+}(\C)$ are unique.
 See for example \cite{BP}.

 A $\phi$ is called trace preserving if $\phi$ is cone
 preserving, and $\tr(\phi(X))=\tr(X)$ for all $X\in
 \cS_n(\C)$.  Note that for a trace preserving $\phi$ we have
 $\phi(\cS_{n,+,1})\subset \cS_{m,+,1}$.

 Recall that a linear operator $\tau:\cS_n(\C)\to\cS_m(\C)$ is
 called \emph{completely positive} if (\ref{defcomposop})
 holds.
 In Kronecker notation (\ref{kronprop})
 \begin{equation}\label{kronrepcompos}
 \tau=\sum_{i=1}^l \bar A_i\otimes A_i.
 \end{equation}
 (Note that the complex space $\C^{n\times n}$ is
 $\cS_n(\C)+(\sqrt{-1})\cS_n(\C)$, and $\tau$ is a \emph{real}
 transformation.)
 Observe that if $A_1,\ldots,A_l\in\R^{m\times n}$ then
 $\tau(\cS_n(\R))\subset \cS_m(\R)$.
 Clearly, completely positive operator is cone preserving.
 Furthermore,
 \begin{equation}\label{conjcompos}
 \tau^*(Y)=\sum_{i=1}^l A_i^* Y A_i \textrm{ where } Y\in
 \cS_m(\C).
 \end{equation}
 Observe that
 \begin{equation}\label{tracid}
 \tr \tau(X)=\tr (X\bA'), \quad \bA':=\sum_{i=1}^l A_i^*A_i.
 \end{equation}
 Hence $\tau$ is trace preserving if and only if $\bA'=I_n$.
 We will assume the condition (\ref{tracecon}), unless stated otherwise.
 Such a mapping $\tau$ is called \emph{a quantum channel}.
 \begin{theo}\label{basineq} Let $\tau:\cS_{n}(\C)\to
 \cS_m(\C)$ be a linear transformation.  Then, for all
 $k=1,\ldots,m$, we have
 \begin{equation}\label{basmaxchar}
 \max_{X\in \cS_{n,+,1}} \sum_{j=1}^k \lambda_j(\tau(X))=\max_{\x\in\C^n,
 \y_1,\ldots,\y_k\in\C^m, \x^*\x=1,\y_p^*\y_q=\delta_{pq}} \sum_{j=1}^k
 \tr(\tau(\x\x^*)(\y_j\y_j^*)).
 \end{equation}
 In particular
 \begin{equation}\label{basineq1}
 \max_{X\in \cS_{n,+,1}} \lambda_1(\tau(X))\le \sigma_1(\tau).
 \end{equation}
 Assume furthermore that $\tau$ is completely positive, i.e.
 (\ref{defcomposop}) holds.  Then, for all $k=1,\ldots,m$, we have
 \begin{eqnarray}\label{basmaxchar1}
 \max_{X\in \cS_{n,+,1}} \sum_{j=1}^k \lambda_j(\tau(X))=\\
 \max_{\x\in\C^n,
 \y_1,\ldots,\y_k\in\C^m, \x^*\x=1,\y_p^*\y_q=\delta_{pq}}
 \sum_{i,j=1}^{l,k}
 | \y_j^* A_i\x|^2.
 \nonumber
 \end{eqnarray}
 In particular,
 \begin{equation}\label{basmaxchar2}
 \max_{X\in \cS_{n,+,1}} \sum_{j=1}^k \lambda_j(\tau(X))\le
 \sum_{j=1}^k \lambda_j(\bA(\tau)), \quad j=1,\ldots,m,
 \end{equation}
 where $\bA(\tau)$ is given by (\ref{defboldA}).

 \end{theo}
 \proof  (\ref{kyfan}) yields that
 $\sum_{j=1}^k \lambda_j(Y)$
 is a convex function on $\cS_m(\C)$, e.g. \cite{Fri81a}.
 Therefore, $\sum_{j=1}^k \lambda_i(\tau(X))$ is a convex function
 on $\cS_{n,+,1}$.  Since the extreme points of $\cS_{n,+,1}$
 are $\x\x^*, \x\in\C^n, \x^*\x=1$, we obtain
 $$\max_{X\in \cS_{n,+,1}} \sum_{j=1}^k
 \lambda_j(\tau(X))=\max_{\x\in\C^n,\x^*\x=1} \sum_{j=1}^k
 \lambda_j(\tau(\x\x^*)).$$
 Combine this equality with (\ref{kyfan}) to deduce
 (\ref{basmaxchar}).
 Compare the maximum characterization (\ref{maxchar}) of
 $\sigma_1(\tau)$ with (\ref{basmaxchar}), ($k=1$), to deduce
 (\ref{basineq1}).

 Assume now that (\ref{defcomposop}) holds.  Note that
 $$\tr((A_i\x\x^*A_i^*)\y_j\y_j^*)=\tr
 ((\y_j^*A_i\x)(\x^*A_i^*\y_j))=|\y_j^*A_i\x|^2.$$
 Hence, for completely positive operator (\ref{basmaxchar})
 is equivalent to (\ref{basmaxchar1}).  The Cauchy-Schwarz
 inequality yields
 $$|\y_j^*A_i\x|^2=|(A_i^*\y_j)^*\x|^2\le
 \|A_i^*\y_j\|^2\|\x\|^2=\y_j^* A_iA_i^*\y_j.$$
 Hence, the left-hand side of (\ref{basmaxchar1}) is bounded
 above by
 $$\max_{\y_1,\ldots,\y_k\in\C^n,\y_p^*\y_q=\delta_{pq}}
 \sum_{j=1}^k \y_j^*\bA(\tau)\y_j.$$
 (\ref{kyfan}) yield that the above maximum is equal to
 $\sum_{j=1}^k \lambda_j(\bA(\tau))$, which implies
 (\ref{basmaxchar2}).  \qed

 \section{Lower bounds on minimal entropies}
 Recall that minimum entropy output of a quantum channel $\tau$,
 denoted by $\rH(\tau)$, is defined by (\ref{defminentrop}).
 Since $\rH(Y)$ is a concave function on $\cS_{m,+}(\F)$,
 and the extreme points of $\cS_{n,+}(\F)$ are of the form
 $\x \x^*$, where $\x\in\F^n$ and $\x^*\x=1$ it follows that
 \begin{equation}\label{minentropchar}
 \rH(\tau)=\min_{\x\in\C^n, \x^*\x=1} \rH(\tau(\x\x^*)).
 \end{equation}
 Assume $\tau_j:\cS_{n_j}(\C)\to\cS_{m_j}(\C), j=1,2$ are two quantum
 channels:
 \begin{equation}\label{2qunchan}
 \tau_j(X_j)=\sum_{i=1}^{l_j} A_{i,j}X_j A_{i,j}^*, \;
 A_{i,j}\in \C^{m_j\times n_j},\;i=1,\ldots,l_j,\;j=1,2.
 \end{equation}
 I.e.
 $$\tau_j=\sum_{i_j=1}^{l_j} \bar A_{i_j,j}\otimes A_{i_j,j},
 \quad j=1,2.$$
 Then $\tau_1\otimes\tau_2$ is quantum channel since
 \begin{equation}\label{tenprodqc}
 \tau_1\otimes\tau_2=\sum_{i_1=i_2=}^{l_1,l_2}
 (\bar A_{i_1,1}\otimes \bar A_{i_2})\otimes (A_{i_1,1}\otimes
 A_{i_2}).
 \end{equation}
 Also, it is straightforward to check that
 \begin{equation}\label{Atau12ident1}
 \bA(\tau_1\otimes\tau_2)=\bA(\tau_1)\otimes\bA(\tau_2).
 \end{equation}
 Hence
 \begin{equation}\label{Atau12ident2}
 \log\lambda_1(\bA(\tau_1\otimes\tau_2))=\log\lambda_1(\bA(\tau_1))+\log\lambda_1(\bA(\tau_2)).
 \end{equation}
 Thus $\log\lambda_1(\bA(\tau))$ is the first additive invariant on quantum
 channels.
 Note that
 $$\cS_{n_1}(\C)\otimes \cS_{n_2}(\C)\subset \cS_{n_1n_2}(\C),\quad
 \cS_{n_1,+,1}(\C)\otimes \cS_{n_2,+,1}(\C)\subset
 \cS_{n_1n_2,+,1}(\C).$$
 Hence we obtain that the minimum entropy output of quantum
 channels is subadditive (\ref{subadit}).
 The additivity conjecture in quantum information theory states
 that equality always holds in (\ref{subadit}) \cite{Sho03}.

 Let $\tau:\cS_n(\C)\to \cS_m(\C)$ be a quantum channel.
 Then the sequence $\rH(\otimes^p \tau)$ is subadditive:
 $$\rH(\otimes^{p+q}\tau)\le
 \rH(\otimes^p\tau)+\rH(\otimes^q\tau) \textrm{ for all integers
 } p,q\ge 1.$$
 Hence the limit (\ref{defregminent}) exists.

 The aim of this paper to give a nontrivial lower bound on
 $\rH_r(\tau)$ for certain quantum channels.
 Assume that $\tau_1,\tau_2$ are two quantum channels given by
 (\ref{2qunchan}).  Viewing $\tau_1,\tau_2$ as linear
 transformation we get
 \begin{equation}\label{logadsvqc}
 \log\|\tau_1\otimes\tau_2\|=\log \sigma_1(\tau_1\otimes\tau_2)=
 \log\sigma_1(\tau_1)+\log\sigma_1(\tau_2)=\log\|\tau_1\|+\log\|\tau_2\|.
 \end{equation}
 Hence, $\log\|\tau\|$ is the second additive invariant on quantum
 channels.
 \begin{theo}\label{mthm1}  Let $\tau:\cS_n(\C)\to\cS_m(\C)$ be
 a quantum channel.  Assume that $\min(\lambda_1(\bA(\tau)),\|\tau\|)<1$.  Then
 \begin{equation}\label{mlowbnd1}
 \rH_r(\tau)\ge \max(-\log\lambda_1(\bA(\tau)), -\log\|\tau\|).
 \end{equation}
 \end{theo}
 \proof  Let $Y\in \cS_{m,+,1}(\C)$.  Since
 $\lambda_1(Y)\ge\ldots\ge\lambda_m(Y)\ge 0$
 $$
 \rH(Y)=\sum_{i=1}^m \lambda_i(Y)\log\frac{1}{\lambda_i(Y)}\ge
 \sum_{i=1}^m \lambda_i(Y)\log\frac{1}{\lambda_1(Y)}\ge
 -\log\lambda_1(Y).$$
 (\ref{basmaxchar2}) for $k=1$, (\ref{Atau12ident1}) and (\ref{basineq1}) yield
 \begin{eqnarray*}
 \rH(\otimes^p\tau)\ge
 -\log\lambda_1(\bA(\otimes^p\tau))=-\log\lambda_1(\otimes^p\bA(\tau))=-p\log\lambda_1(\bA(\tau)),\\
 \rH(\otimes^p\tau)\ge
 -\log\sigma_1(\otimes^p\tau)=-p\log\sigma_1(\tau)=-p\log\|\tau\|
 \end{eqnarray*}
 Hence (\ref{mlowbnd1}) holds.  \qed

 Note that the proof of the above theorem yields that (\ref{mlowbnd1}) always
 holds.  However if $\min(\lambda_1(\bA(\tau)),\|\tau\|)\ge 1$ then the inequality (\ref{mlowbnd1})
 is trivial.
 \begin{prop}\label{lowbdl1a}  Let $\tau$ be a quantum channel
 given by (\ref{defcomposop}).  Then
 \begin{equation}\label{lowbdl1b}
 \lambda_1(\bA(\tau))\ge \frac{n}{m},\quad \sigma_1(\tau)\ge
 \frac{\sqrt{n}}{\sqrt{m}}.
\end{equation}
 Hence, $\lambda_1(\bA(\tau)),\sigma_1(\tau)\ge 1$ for $m\le
 n$.  In particular, if $m\le n$ then the condition either $\lambda_1(\bA(\tau))=1$
 or $\sigma_1(\tau)=1$ holds if and only if
 $m=n$ and $\tau^*$ is a quantum channel.
 \end{prop}
 \proof Clearly,
 $$m\lambda_1(\bA(\tau))\ge\sum_{j=1}^m \lambda_j(\bA(\tau))=\tr
 \bA(\tau)=\sum_{i=1}^l
 \tr A_i A_i^*=\sum_{i=1}^l \tr A_i^* A_i=\tr I_n=n.$$
 Hence $\lambda_1(\bA(\tau))\ge \frac{n}{m}$.  Clearly, if
 $m=n$ and $\bA(\tau)=I_n$ then $\lambda_1(\bA(\tau))=1$ and
 $\tau^*$ is a quantum channel.  Vice versa if $m\le n$ and
 $\lambda_1(\bA(\tau))=1$ then $m=n$.  Furthermore, all eigenvalues of $\bA(\tau)$
 have to be equal to $1$, i.e. $\bA(\tau)=I_n$.

 Observe that the condition that $\tau$ of the form
 (\ref{defcomposop}) is a quantum channel is equivalent to the
 condition $\tau^*(I_m)=I_n$.  As
 $$\sigma_1(\tau)=\sigma_1(\tau^*)\ge
 \|\tau^*(\frac{1}{\sqrt{m}} I_m)\|=\frac{\sqrt{n}}{\sqrt{m}}$$
 we deduce that second inequality in (\ref{lowbdl1b}).
 Suppose that $m\le n$ and $\sigma_1(\tau)=1$.  Hence $m=n$ and
 $\sigma_1(\tau^*)=\|\tau^*(\frac{1}{\sqrt{n}}
 I_n)\|=1$.  So $\frac{1}{\sqrt{n}}I_n$ must be the left and
 the right singular vector of $\tau$ corresponding to the
 $\|\tau\|$.  I.e. $\tau(I_n)=I_n$, which is equivalent to the
 condition that $\tau^*$ is a quantum channel.
 \qed

 In the next sections we will give examples for which
 $\lambda_1(\bA(\tau))<1$.   In that case we can
 improve the lower bound for $\rH_r(\tau)\ge -\log \lambda_1(\bA(\tau))$.
 Denote by $m'\ge 1$ the smallest positive integer that
 \begin{equation}\label{defm'}
 \sum_{i=1}^{m'}\lambda_i(\bA)\ge 1.
 \end{equation}
 Since $\tau$ is trace preserving (\ref{basmaxchar2}) yields
 that $m'\ge m$.  Note that $m'>1$ if and only if
 $\lambda_1(\bA(\tau))<1$.  Assume first that $m'>1$.  Let
 \begin{eqnarray}\label{defetA}
 \rF(\bA(\tau))=-\eta(\bA(\tau))\log\eta(\bA(\tau))-\sum_{i=1}^{m'-1}
 \lambda_i(\bA(\tau))\log\lambda_i(\bA(\tau)),\\
 \textrm{ where }\eta(\bA(\tau))=1-\sum_{i=1}^{m'-1} \lambda_i(\bA(\tau)).\nonumber
 \end{eqnarray}
 Note that in this case $0\le\eta(\bA(\tau))\le
 \lambda_{m'}(\bA(\tau))$.  Hence
 \begin{equation}\label{ineqFlmb}
 F(\bA(\tau))\ge -\log\lambda_1(\bA(\tau)).
 \end{equation}
 If $\lambda_1(\bA(\tau))\ge 1$ we let $\rF(\bA(\tau))=0$.
 \begin{theo} Let $\tau$ be a quantum channel given by
 (\ref{defcomposop}).  Let $\bA(\tau)$ be given by (\ref{defboldA})
 and assume that $\rF(\bA(\tau))$ is defined as above.  Then
 \begin{equation}\label{lowbdbA}
 \rH_r(\tau)\ge \limsup_{p\to\infty}\frac{\rF(\otimes^p\bA(\tau))}{p}.
 \end{equation}
 \end{theo}
 \proof If $\lambda_1(\bA)\ge 1$ then
 $\lambda_1(\otimes^p\bA(\tau))=\lambda_1(\bA)^p\ge 1$ and
 $F(\otimes^p\bA(\tau))=0$.  In that case (\ref{lowbdbA}) is
 trivial.

 Assume that $\lambda_1(\bA(\tau))<1$.  Let
 $$\et(\bA(\tau)):=(\lambda_1(\bA(\tau)),\ldots,\lambda_{m'-1}(\bA(\tau)),
 \eta(\bA(\tau)),\underbrace{0,\ldots,0}_{m-m'})\trans\in \R_+^m.$$
 (\ref{basmaxchar2}) implies that $\lam(\tau(X))\prec
 \et(\bA(\tau))$ for each $X\in\cS_{n,+,1}$.  Since $x\log x$
 is convex on $\R_+$ it follows that $-\rH(\tau(X))\le
 -\rF(\bA(\tau))$.  Hence $\rH(\tau)\ge \rF(\bA(\tau)$.
 Similarly
 $$\rH(\otimes^p\tau)\ge \rF(\bA(\otimes^p\tau))=\rF(\otimes^p
 \bA(\tau)).$$
 Hence (\ref{lowbdbA}) holds in this case.  \qed

 We remark that the inequality (\ref{ineqFlmb}) shows that
 (\ref{lowbdbA}) is an improvement of the inequality
 $\rH_r(\tau)\ge -\log\lambda_1(\bA(\tau))$ when
 $\lambda_1(\bA(\tau))<1$.  Since the eigenvalues of
 $\otimes^p\bA(\tau)$ are rearranged coordinates of the vector
 $\otimes^p \lam(\bA(\tau))$, it should not be too difficult to
 find the exact formula of the right-hand side of
 (\ref{lowbdbA}) in terms of $\lam(\bA(\tau))$.

 \section{Examples}
 \textbf{Example 1}.  A quantum channel
 $\tau:\cS_1(\C)\to\cS_m(\C)$ is of the form
 \begin{equation}\label{1mquntchan}
 \tau(x)=\sum_{i=1}^l \a_i x\a_i^*, \quad \a_i\in \C^m, i=1,\ldots,l,
 \;\sum_{i=1}^l \a_i^*\a_i=1, \quad \bA(\tau)=\sum_{i=1}^l
 \a_i\a_i^*.
 \end{equation}
 Note that $\tr \bA(\tau)=1$.  Hence $\lambda_1(\bA(\tau))<1$,
 unless $\a_1,\ldots,\a_l$ are colinear.  (This happens always if
 $m=1$.)

 We claim that
 \begin{equation}\label{1mquntchansg1}
 \sigma_1(\tau)=\sqrt{\tr \bA(\tau)^2}.
 \end{equation}
 Indeed
 $$\max_{|x|=1,Y\in \cS_m(\C), \tr(Y^2)=1}|\tr \tau(x)Y|=\max_{Y\in \cS_m(\C),
 \tr(Y^2)=1} |\tr \bA(\tau)Y|=\sqrt{\tr \bA(\tau)^2}.$$
 Hence
 \begin{equation}\label{inlamsig}
 \lambda_1(\bA(\tau))<\sigma_1(\tau) <1 \textrm{ iff }
 \a_1,\ldots,\a_l \textrm{ are not colinear}.
 \end{equation}
 If $\a_1,\ldots,\a_l$ are co-linear then
 $\lambda_1(\bA)=\sigma_1(\bA)=1$.
 Note that in this example $\rH(\tau)=\rH(\bA(\tau))$.\\

 \noindent
 \textbf{Example 2}.  A quantum channel
 $\tau:\cS_n(\C)\to\cS_1(\C)$ is of the form
 \begin{equation}\label{n1quntchan}
 \tau(X)=\sum_{i=1}^l \a_i^* X\a_i, \quad \a_i\in \C^n, i=1,\ldots,l,
 \;\sum_{i=1}^l \a_i\a_i^*=I_n, \quad \bA(\tau)=\sum_{i=1}^l
 \a_i^*\a_i=n.
 \end{equation}
 So $\lambda_1(\bA(\tau))=n\ge 1$.  On the other hand
 \begin{equation}\label{n1quntchansg1}
 \sigma_1(\tau)=\max_{X\in\cS_n(\C), \tr
 X^2=1,|y|=1}|\tr(\tau(X)y)|=\max_{X\in\cS_n(\C), \tr X^2=1} |\tr
 X|=\sqrt{n}.
 \end{equation}
 So for $n>1$ $\lambda_1(\bA(\tau))>\sigma_1(\tau)$.\\

 \noindent
 \textbf{Example 3}.  A quantum channel of the form
 (\ref{defcomposop}), where $m=n$ and (\ref{tracecon}) holds, is called a
 strongly self-adjoint if there exists a permutation $\pi$ on
 $\{1,\ldots,l\}$ such that $A_i^*=A_{\pi(i)}$ for
 $i=1,\ldots,l$.
 So $\bA(\tau)=I_n$ and $\lambda_1(\bA(\tau))=1$.  Note that $\tau$
 is self-adjoint and $\tau(I_n)=I_n$.  Since $I_n$ is an interior
 point of $\cS_{n,+}$ it follows that $\sigma_1(\tau)=1$.\\

 \noindent
 \textbf{Example 4}.  Assume $\tau_j:\cS_{n_j}(\C)\to\cS_{m_j}(\C), j=1,2$ are two quantum
 channels.  Consider the quantum channel
 $\tau=\tau_1\otimes\tau_2$.  Then
 $$\log\lambda_1(\bA(\tau))=\log\lambda_1(\bA(\tau_1))+\log\lambda_1(\bA(\tau_2)),\;
 \log\sigma_1(\tau)=\log\sigma_1(\tau_1)+\log\sigma_1(\tau_2).$$
 Thus, it is possible to have $\lambda_1(\bA(\tau))<1$  without the assumption
 that both $\tau_1$ and $\tau_2$ satisfy the same condition.
 Combine Example 1 and Example 3 to obtain examples
 of quantum channels $\tau:\cS_{n}(\C)\to \cS_{mn}(\C)$, where
 $n,m>1$ where $\lambda_1(\bA(\tau))<1$.  Similar arguments
 apply for $\sigma_1(\tau)$.\\

 \noindent
 \textbf{Example 5}.
 Recall that if $B\in \C^{m\times n}$ and $C\in \C^{p\times q}$
 then
 $$B\oplus C=\left[\begin{array}{ll}B&0_{m\times q}\\0_{p\times
 n}&C\end{array}\right]\in \C^{(m+p)\times (n+q)}.$$
 Assume $\tau_j:\cS_{n_j}(\C)\to\cS_{m_j}(\C), j=1,2$ are two quantum
 channels given as in (\ref{2qunchan}).  Then
 $\tau_1\oplus\tau_2: \cS_{n_1+n_2}(\C):\to\cS_{m_1+m_2}(\C)$
 is defined as follows.
 $$(\tau_1\oplus\tau_2)(X)=\sum_{i_1=i_2=1}^{l_1,l_2}
 (A_{i_1,1}\oplus A_{i_2,2}) X (A_{i_1}^*\oplus A_{i_2,2}^*).$$
 Clearly, $\tau_1\oplus\tau_2$ is a quantum channel.
 Furthermore,
 $$\bA(\tau_1\oplus\tau_2)=\bA(\tau_1)\oplus\bA(\tau_2).$$
 Hence
 \begin{equation}\label{lambt1opt2}
 \lambda_1(\bA(\tau_1\oplus\tau_2))=\max(\lambda_1(\bA(\tau_1)),\lambda_1(\bA(\tau_2))).
 \end{equation}
 This if $\lambda_1(\bA(\tau_i))<1$ we get that
 $\lambda_1(\bA(\tau_1\oplus\tau_2)<1$.

 The formula for $\sigma_1(\tau_1\oplus\tau_2)$ does not seems
 to be as simple as (\ref{lambt1opt2}).  By viewing
 $\cS_{n_1}(\C)\oplus\cS_{n_2}(\C)$ as a subspace of
 $\cS_{n_1+n_2}(\C)$ we deduce the inequality
 $$\sigma_1(\tau_1\oplus\tau_2)\ge
 \max(\sigma_1(\tau_1),\sigma_1(\tau_2)).$$
 \textbf{Example 6}.  We first show how to take a neighborhood
 of a given quantum channel given by (\ref{defcomposop}).
 View $\cA:=(A_1,\ldots,A_l)$ as a point in $(\C^{m\times n})^l$.
 Let $\rO(\cA)\subset (\C^{m\times n})^l$ be an open
 neighborhood of $\cA$ such that for any
 $\cB:=(B_1,\ldots,B_l)\in (\C^{m\times n})^l$ the matrix
 $C(\cB):=\sum_{i=1}^l B_i^*B_i$ has positive eigenvalues.
 Define
 $$\hat \cB=(\hat B_1,\ldots,\hat
 B_l)=(B_1C(\cB)^{-\frac{1}{2}},\ldots,B_lC(\cB)^{-\frac{1}{2}})\in (\C^{m\times
 n})^l.$$
 Then $\tau_{\cB}:\cS_{n}(\C)\to \cS_m(\C)$ given by
 $$\tau_{\cB}(X)=\sum_{i=1}^l \hat B_i X (\hat B_i)^*$$
 is a quantum channel.  So if $O(\cA)$ is a small neighborhood
 $\cA$ then $\tau_{\cB}$ is in the small neighborhood of
 $\tau$.   In particular of $\lambda_1(\bA(\tau))<1$ then
 there exists a small neighborhood $O(\cA)$ such that
 $\lambda_1(\bA(\tau_{\cB}))<1$ for each $\cB\in O(\cA)$.
 Similar claim holds if $\sigma_1(\tau)<1$.

 \section{Bi-quantum channels}\label{sec:unqunch}
 \begin{theo}\label{entuqc}  Let $\tau:\cS_n(\C)\to \cS_n(\C)$
 be a bi-quantum channel.  Then $\sigma_1(\tau)=1$. Assume
 that $n\ge 2$ and $\sigma_2(\tau)<1$.  Then
 \begin{equation}\label{entuqc1}
 \rH(\tau)\ge -\frac{1}{2} \log (\sigma_2(\tau)^2 +
 \frac{1-\sigma_2(\tau)^2}{n}).
 \end{equation}
 \end{theo}
 \proof  Observe first that since $\tau$ and $\tau^*$ are
 quantum channels if follows that $\omega:=\tau^*\tau$ is a
 self-adjoint quantum channel on $\cS_n(\C)$.  As $\omega$
 preserves the cone of positive hermitian matrices,
 $\omega(I_n)=I_n$ and $I_n$ is an interior point of
 $\cS_{n,+}(\C)$, the Krein-Milman theorem, e.g. \cite{BP},
 it follows that $1$ is the maximal eigenvalue of $\omega$.
 Hence $\sigma_1(\tau)=1$.  Observe next
 $$\lambda_1(\tau(\x\x^*))\le (\sum_{i=1}^n \lambda_i(\tau(\x\x^*))^2)^{\frac{1}{2}}
 =\|\tau(\x\x^*)\|.$$
 We now estimate $\|\tau(\x\x^*)\|$ from above, assuming that
 $\|\x\|=1$.  Consider the singular value decomposition of
 $\tau$ given by (\ref{singvalcon}-\ref{singvalcon1}).
 Here $m=n$ and we can assume that $U_1=V_1=\frac{1}{\sqrt{n}}
 I_n$.  (\ref{SVDnormid}) yields that
 $$\sum_{i=1}^n \lambda_i(\tau(\x\x^*))^2=\sum_{i=1}^{\rank \tau} \sigma_i(\tau)^2 |\tr
 U_i\x\x^* |^2\le \sigma_1(\tau)^2 |\tr U_1 \x\x^*|^2 +
 \sum_{i=2}^{\rank \tau} \sigma_2(\tau)^2 |\tr
 U_i\x\x^* |^2.$$
 Since $\sigma_1(\tau)=1$ and $\tr U_1\x\x^*=\frac{1}{\sqrt n} \tr \x\x^*=\frac{1}{\sqrt
 n}$, we deduce that
 \begin{equation}\label{basinuc}
 \sum_{i=1}^n \lambda_i(\tau(\x\x^*))^2\le \sigma_2(\tau)^2+
 \frac{1-\sigma_2(\tau)^2}{n}.
 \end{equation}
 So
 $$\lambda_1(\tau(\x\x^*)\le \sqrt{\sigma_2(\tau)^2+
 \frac{1-\sigma_2(\tau)^2}{n}}.$$
 Use the arguments of the proof of Theorem \ref{mthm1} to
 deduce (\ref{entuqc1}).  \qed
 \begin{prop}\label{tproduch}  Let $\tau_i:\cS_{n_i}(\C)\to \cS_{n_i}(\C)$ be
 a bi-quantum channel for $i=1,2$.  Then $\tau_1\otimes\tau_2$ is a
 bi-channel.  Furthermore
 \begin{equation}\label{tproduch1}
 \sigma_2(\tau_1\otimes\tau_2)=\max(\sigma_2(\tau_1),\sigma_2(\tau_2)).
 \end{equation}
 In particular, if $\tau:\cS_n(\C)\to \cS_n(\C)$ is a unitary channel and
 $\sigma_2(\tau)<1$ then
 \begin{equation}\label{tproduch2}
 \rH(\otimes^p\tau)\ge -\frac{1}{2} \log (\sigma_2(\tau)^2 +
 \frac{1-\sigma_2(\tau)^2}{n^p}).
 \end{equation}
 \end{prop}
 \proof  Since
 $(\tau_1\otimes\tau_2)^*=\tau_1^*\otimes\tau_2^*$ it follows
 that a tensor product of two
 bi-quantum channels is a bi-quantum channel.
 Since the singular values of $\tau_1\otimes \tau_2$ are
 all possible products of of singular values of $\tau_1$ and
 $\tau_2$ we deduce (\ref{tproduch1}).  Then (\ref{tproduch2})
 is implied by Theorem \ref{entuqc}.  \qed
 \begin{lemma}\label{sigeles1}  Consider a unitary channel of
 the form (\ref{defcomposop}) and (\ref{spectau}), where $l\ge
 3$, $t_i\ne 0, i=1,\ldots,l$, $Q_1=I_n$, and $Q_2,\ldots,Q_{l}$
 do not have a common nontrivial invariant subspace.  Then
 $\sigma_2(\tau)<\sigma_1(\tau)=1$.
 \end{lemma}
 \proof  Assume that $X\in
 \cS_{n,+}(\C)$ has rank $k\in [1,n-1]$. We claim that $\rank
 \tau(X)>\rank X$.  Recall that $X=\sum_{j=1}^k \x_j\x_j^*$,
 where $\x_1,\ldots, \x_k\in\C^n$ are nonzero orthogonal
 vectors.  As $t_1^2,\ldots,t_k^2>0$ we deduce that
 $$\tau(X)=t_1^2 X+\sum_{j=2}^k t_j^2 Q_jXQ_j^*\ge t_1^2 X.$$
 So $\rank\tau (X) \ge k$.  Furthermore $\rank\tau(X)=k$
 if and only $Q_i\x_j\in\U:=\span(\x_1,\ldots,\x_k)$ for
 $i=2,\ldots,l$ and $j=1,\ldots,k$.  Since $\U$ is not
 invariant under $Q_2,\ldots,Q_l$ we deduce that
 $\rank\tau(X)>k$.  Clearly, if $Y\ge 0$ and $\rank Y=n$ then
 $\rank\tau(Y)=n$.

 Observe next that $Q_2^*,\ldots,Q_l^*$ do not have a
 nontrivial common invariant subspace.  Indeed, if
 $\V\subset\C^n$ was a nontrivial common invariant of
 $Q_2^*,\ldots,Q_l^*$, then the orthogonal complement of $\V$
 will be a nontrivial invariant subspace of $Q_2,\ldots,Q_l$,
 which contradicts our assumption.  Hence $\tau^*(X)>\rank X$.

 Let $\eta=\tau^*\tau$.
 Thus, $\rank \eta^n(Z) =n$ for any $Z\gneq 0$,
 i.e., $\eta^n$ maps $\cS_{n,+}(\C)\backslash\{0\}$ to
 the interior of  $\cS_{n,+}(\C)$.  By Krein-Milman theorem,
 i.e. \cite{BP}, $1=\lambda_1(\eta^n)>\lambda_2(\eta^n)=\sigma_2(\tau)^{2n}$.
 \qed
 \begin{corol}  Let $\tau:\cS_n(\C)\to \cS_n(\C)$ be a generic
 unitary quantum channel.  I.e. $\tau$ of
 the form (\ref{defcomposop}) and (\ref{spectau}), where $l\ge
 3$, $(t_1^2,\ldots,t_l^2)\trans$ is a random probability
 vector, and $Q_1,\ldots,Q_l$ are random unitary matrices.
 Then $\sigma_2(\tau)<\sigma_1(\tau)=1$.
 \end{corol}
 \proof Let $\tau_1(X):=\tau(Q_1^*XQ_1)$.  Clearly, the $l-1$
 unitary matrices $Q_2Q_1^*,\ldots,Q_lQ_1^*$ are $l-1$ random
 unitary matrices.  Since $l-1\ge 2$ these $l-1$ matrices do
 not have a nontrivial common invariant subspace.  Lemma \ref{sigeles1}
 yields that $\sigma_2(\tau_1)<1$.  Clearly,
 $\sigma_2(\tau_1)=\sigma_2(\tau)$.  \qed

 \bibliographystyle{plain}

 \emph{Acknowledgement}:  I thank Gilad Gour
 for useful remarks.

\end{document}